\documentclass[12pt]{article}

\usepackage{amsfonts,amssymb,amsmath,epsfig,latexsym}

\newcommand{\pd}{\partial}
\newcommand{\vect}[1]{\boldsymbol #1}

\DeclareMathOperator{\sign}{sign}

\begin{document}

\title{
Interference as a statistical consequence\\
of conjecture on time quant
}

\author{
Andrei Khrennikov$^\text{\hspace{0.25mm}a}$ and Yaroslav Volovich\footnote{%
Supported in part by the EU Human Potential Programme,
contract N. HPRN-CT-2002-00279 (Network on Quantum Probability and Applications)
and Profile Math. Modelling in Physics and Cogn. Sc. of V\"{a}xj\"{o} University.}\\
~\\
International Center for Mathematical Modeling\\
in Physics, Engineering and Cognitive science\\
MSI, V\"axj\"o University, SE-351 95, Sweden\\
~\\
email: Andrei.Khrennikov@msi.vxu.se\\
email: Yaroslav.Volovich@msi.vxu.se
}

\date{}

\maketitle

\begin{abstract}
We analyze statistical consequences of a conjecture that there
exists a fundamental (indivisible) quant of time. We study
particle dynamics with discrete time. We show that a quantum-like
interference pattern could appear as a statistical effect for
deterministic particles, i.e. particles that have trajectories and
obey deterministic dynamical equations, if one introduces a
discrete time. As a demonstration of this concept we consider
particle scattering on a screen with a slit. We study how
resulting interference picture depends on the parameters of the
model. The resulting interference picture has a nontrivial
minimum-maximum distribution which vanishes, as the time
discreteness parameter goes to zero. This picture is qualitatively
the same as one obtained in quantum experiments. The picture
includes some interesting nonclassical properties such as a
`black' region behind the center of the slit.
\end{abstract}

\section{Introduction}

In papers\cite{KhVol1,KhVol2,KhVolJMO} we presented the conjecture:

\textbf{there exists a fundamental quant of time $\tau$.}

\noindent
In fact we came to this conjecture by finding (by pure occasion) that
statistical histograms obtained for classical systems in the one or two
slit experiments can have the form of quantum-like interference observed in
experiments with photons or electrons. We started to speculate that
essential features of quantum statistics (at least some of them) could
be obtained as a consequence of existence of time quant.

Basically speaking the idea of discretization is quite natural for physics.
Indeed discretization of energy was exploited by Plank-Einstein ideas in order to
explain such quantum phenomena as black body radiation, photo-electric effect, etc.
On the other hand from the point of view of energy-time uncertainty relations
discretization of time gives certain constraints to a (measured) energy of the system.

On the other hand by introducing discretized time one effectively
changes the theoretical structure of the space-time on small
distances\footnote{In particular, this idea was exploited by p-adic
space-time alternative\cite{VVZ,FW,FO,Kh-p-adic}}.
From this point of view it is rather natural to assume that a value
of the discreteness parameter is of the order of Plank time
$t_{Pl}\approx 5\cdot 10^{-44} (sec)$.
One could think of discretized time as a lack of information of where the
particle is during the discretization period.

In this paper we study deterministic model for particle scattering
on a screen with a single slit.
Resulting interference picture has a nontrivial minimum-maximum
distribution which is qualitatively the same as one obtained in
quantum experiments.
The interference pattern appears as a statistical effect of particles
which have trajectories\footnote{Cf. with Bohmian mechanics\cite{Bohm,BH}}.
The interference picture has no relation to
`self-inter\-feren\-ce' of particles, no wave-structure is involved
into considerations. The basic source of interference is the
discrete time scale used in our mathematical model: instead of
Newton's differential equations with continuous time evolution, we
consider difference equations with discrete time evolution.
Interference effect disappears as the time discreteness parameter goes to zero.
The picture includes some interesting nonclassical properties such
as a `black' region behind the center of the slit\footnote{We would
like to thank G.~Emch for an interesting question he asked on the
`International Conference: Reconsideration of Foundations-2' held in V\"axj\"o --
whether it is possible to reproduce Fresnel-like phenomenon of a black
region behind the center of the slit in the discrete time formalism -- our answer
now is \textit{yes} (see section \ref{sec:sslit} for details).}.

The paper is organized as follows. In the next section we provide basic ideas
of the discrete time formalism and compare it with construction of quantum mechanics.
In section \ref{sec:ddyn} we write dynamical equations with discrete time and in
section \ref{sec:newton} we present them in intuitive form.
In section \ref{sec:sslit} we apply the formalism to a particle
scattering on a screen with a slit and study appearance and properties of the resulting
interference picture. Finally, in the appendix we provide a detailed description of
the numerical simulation performed, we list a \textit{Mathematica} program which
mirrors the original parallel \textit{C++} programm which we used to perform
actual simulation.

We remind that recently it was suggested\cite{Kh1,Kh2,Kh3,Kh4,Kh-interp}
that quantum mechanical
interference rules could be explained using contextual transformations.
This approach makes one think of a construction of physically
reasonable model which produces interference resting on the
classical rule of addition of probabilistic alternatives
(extended with a notion of \textit{context}).
Our quantum-time approach could be considered as a candidate for such a model.

\section{Basic Steps of the Approach}
\label{sec:basics}

According to the principles of quantum mechanics in order to
compute measurable quantities, in the simplest case, one has to
perform the following steps\cite{Heis,Heis2}. First, the form of classical and
quantum equations of motion remains the same (we are talking about
Heisinberg's representation), although the
dynamical variables are now noncommutative.
Next, one has to add to the original initial conditions an extra one stating
commutative properties of the canonical variables, like
$[\hat{x}(0),\hat{p}(0)]=i\hbar$.
At the last step one says that measurable quantities are eigenvalues of the
dynamical variables. Here the second step introduces the Plank constant which
distinguishes classical and quantum worlds. The noncommutative
property of dynamical variables is the root of the fact that
quantum particles can not have trajectories and introduction
of quantum state implies the nondeterministic behavior of quantum
systems.

In this note we tried to achieve the same results for measurable
quantities as in quantum mechanics (at least in the part vastly
supported by experiments), but trying to keep trajectories
and deterministic behavior of particles.
We follow steps similar to the above ones,
but we keep the measurable quantities being real functions.
Nevertheless, we have had to introduce the time discreteness parameter $\tau$.
We start from classical equations of motion with discrete time and postulate the
canonical variables to be measurable.

\section{Discrete Time Dynamics}
\label{sec:ddyn}

In both classical and quantum mechanics a dynamical function
$A=A(p,q)$ evolves according to the following well known equation
\begin{equation}
\label{dyneq} D_t A=[H,A]
\end{equation}
where $H=H(p,q)$ is a Hamiltonian of the system and in the right
hand side is a Poisson bracket, either classical or
quantum\footnote{Poisson bracket in classical mechanics could be
presented as
\begin{equation}
\label{cposs}
[A,B]=\frac{\pd A}{\pd p}\frac{\pd B}{\pd q} -
\frac{\pd A}{\pd q}\frac{\pd B}{\pd p}
\end{equation}
and in quantum mechanics it is a commutator
$[A,B]=\frac{1}{i\hbar}(AB-BA)$
}.

In both classical and quantum dynamics the left hand side of
(\ref{dyneq}) is the same, it contains a continuous time
derivative
$$
D_t A = \frac{dA}{dt}
$$

As it was mentioned in the previous section we are interested in
construction dynamics with discrete time. This is done with the
help of \textit{discrete derivative} which is postulated to be
$$
D^{(\tau)}_t A = \frac{1}{\tau}[A(t+\tau)-A(t)],
$$
where $\tau$ is the discreteness parameter. This parameter is
finite and is treated in the same way as Plank constant in quantum
mechanical formalism. In particular if $\tau$ is small relative to
dimentions of the system then classical approximation with
continuous derivative should work well (although this could not be
the case all the time in the same sense as there are examples when
quantum formalism is reasonable even for macroscopic systems, for
example in superfluidity).

Thus the discrete time dynamical equation is postulated to be
$$
D^{(\tau)}_t A=[H,A],
$$
where $A(p,q)$ is a real-valued function of real-valued dynamical
variables and in the right hand side there is the classical
Poisson bracket (see (\ref{cposs})).

Please note that in our model the coordinate space is left
continuous.

\section{Discrete Time in Newton's Equations}
\label{sec:newton}

In this section we provide an intuitive description of the
discretization procedure described above.
Consider a well known Newton's equation
\begin{equation}
\label{eq} \vect{F}=m\ddot{\vect{r}}
\end{equation}
This equation gives the same dynamics as
(\ref{dyneq})-(\ref{cposs}).
Let us now modify this equation, introducing the time discreteness
parameter $\tau$.
We rewrite the second order differential equation (\ref{eq})
as a system of first order differential equations, we have
\begin{equation}
\label{e1}
\begin{split}
\vect{F}&=m\dot{\vect{v}}\\
\vect{v}&=\dot{\vect{r}}
\end{split}
\end{equation}
In the system (\ref{e1}) the derivatives assume continuousness of time.
Let us now introduce the discreteness parameter $\tau$. We have
\begin{equation}
\label{e2}
\begin{split}
\vect{F}&=m \frac{\vect{v}(t+\tau)-\vect{v}(t)}{\tau}\\
\vect{v}(t+\tau)&=\frac{\vect{r}(t+\tau)-\vect{r}(t)}{\tau}
\end{split}
\end{equation}
In the limit of $\tau\to 0$ (\ref{e2}) is equivalent to (\ref{eq})
and (\ref{e1}).

Now let us rewrite the system (\ref{e2}) in a directly computable
way
\begin{equation}
\begin{split}
\vect{v}(t+\tau)&=\vect{v}(t)+\vect{F}\tau/m\\
\vect{r}(t+\tau)&=\vect{r}(t)+\vect{v}(t)\tau
\end{split}
\end{equation}
where $\vect{F}=\vect{F}(\vect{r}(t),\vect{v}(t),t)$.

\section{Interference Pattern in a Single Slit Scattering}
\label{sec:sslit}

\begin{figure}[ht]
\begin{center}
\epsfig{file=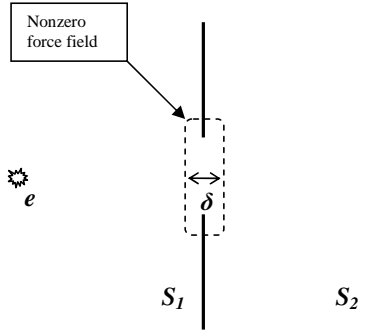,width=5cm}
\caption{Particle scattering on a screen with a slit.
Particles are emitted from the source e pass through a slit in screen
$S_1$ and gather on screen $S_2$.
There is nonzero force field in the region $\delta$ near the slit.}
\label{fig:setup}
\end{center}
\end{figure}

Consider the following experimental setup (see Fig.\ref{fig:setup}).
A particle source $e$ is located in front of the center of a slit in a screen $S_1$.
Near the slit there is a nonzero force field which affects particles,
$$
\vect{F}(x,y)\neq 0 \mbox{,~~for~} (x,y)\in\delta
$$
($x$ and $y$ are coordinates along horizontal and vertical axes respectively,
the axes origin is in the center of the slit).
Particles pass through the slit and concentrate on a second screen $S_2$.
We study a particle density on the screen $S_2$.

\begin{figure}[ht]
\begin{center}
\epsfig{file=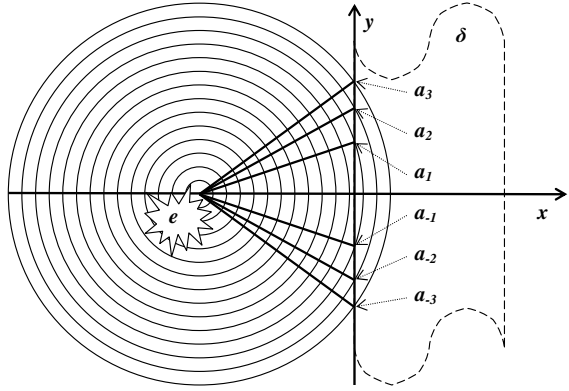,width=7.5cm}
\caption{Particles emitted from the source $e$ enter the region $\delta$.
Points $a_{\pm1}$, $a_{\pm2}$, $a_{\pm3}$, \ldots are origins of deviation which
forms interference pattern.}
\label{fig:circs}
\end{center}
\end{figure}

\begin{figure}
\centering
\epsfig{width=5cm,file=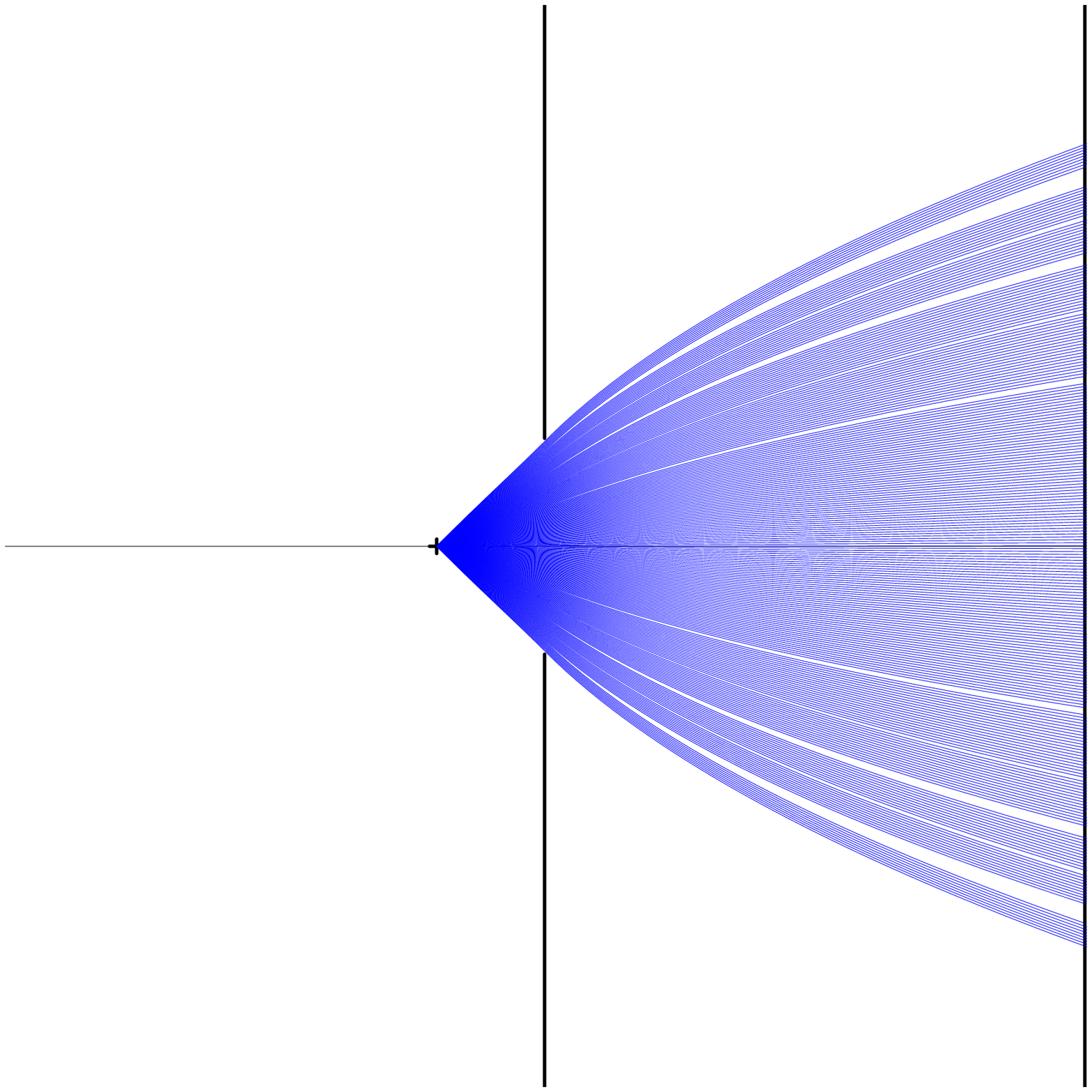}
\epsfig{width=5cm,file=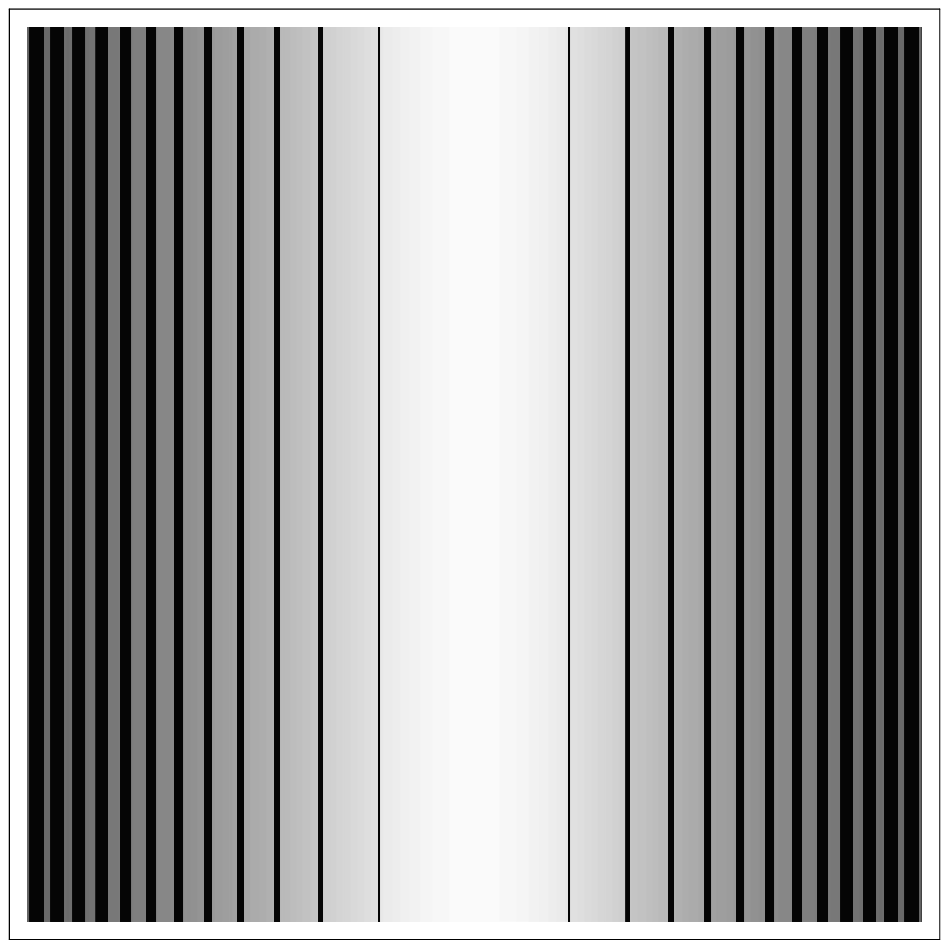}
\caption{Interference pattern (brighter area corresponds to the higher particle density) and
particle trajectories in a single-slit scattering computed in a discrete time formalism.}
\label{fig:histg}
\end{figure}

Let us start from the simplest case when the force field is constant and perpendicular
to the screens, $\vect{F}(x,y)=\vect{e}_x~F_0$, for $x\geqslant 0$ and $0$ otherwise.
Let the source $e$ be point-like emitting particles with constant velocity $v_0$
under random evenly distributed angles $\alpha\in[0,2\pi)$.
In this case trajectories of particles emitted by $e$ (in discrete time dynamics)
in the region $x<0$ (i.e. before the screen $S_1$) form concentric circles originating
from $e$ with the radii $r_n=v_0\tau n$, $n=1,2,\ldots$ (Fig.\ref{fig:circs}).
We get the circles -- which are trajectories of many particles emitted under close
angles -- in the region where there is no force field and particles move along strait
lines, exactly as in classical dynamics.
Let $a_i$ be the points where these circles enter the region $\delta$.
The distance between the center of the slit and $a_i$ is given by
\begin{equation}
\label{ai}
a_i=\sign(i)\sqrt{v_0^2\tau^2(n_0+|i|)^2-d^2},
\end{equation}
where $d$ is a distance between $e$ and $S_1$, and
$n_0$ is the largest integral value not greater than a fraction $d/v_0\tau$,
\begin{equation}
\label{n0}
n_0=\left\lfloor \frac{d}{v_0\tau}\right\rfloor
\end{equation}
This rather simple setup already produces interesting nontrivial interference pattern
(Fig.\ref{fig:histg}).

The points $a_i$ are the origins of deviation from `classical' (i.e. continuous time)
trajectories. This deviation forms the interference pattern.
To argument this let $\varphi_i$ be an angle between the horizontal axis and a
line connecting $e$ and $a_i$. Particles emitted under angles less than $\varphi_i$
($i\geqslant 1$) become affected by the force field in the region $\delta$ one step
earlier than those emitted under angles grater than $\varphi_i$.
As a result there appears a `fork' -- even very close trajectories but with the
angles above and below $\varphi_i$ become separated.
One could get the points of minima of the interference pattern by following
the trajectory along the line $e\rightarrow a_i$ and the parabolic curve
(movement in a constant field) in the region $x>0$ until the trajectory hits the screen
$S_2$.

The case described above is not completely physically reasonable, indeed
is requires a force field everywhere in the half-plane $x\geqslant 0$,
thus it could be considered only as a simplified model which still demonstrates
some interesting properties of discrete time dynamics.
An example of such a property is a Fresnel-like phenomenon of a black
region behind the center of the slit which is discussed below.

\begin{figure}
\centering
\epsfig{width=5.5cm,file=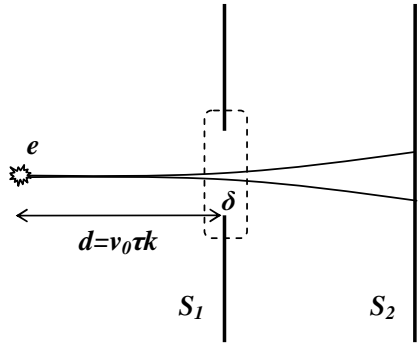}
\caption{Fresnel-like phenomenon of a black
region behind the center of the slit in the discrete time formalism.}
\label{fig:black}
\end{figure}

One could see from (\ref{ai})-(\ref{n0}) that if
$$
d=v_0\tau k, \mbox{,~~where~}k=1,2,\ldots
$$
then $a_{+1}=a_{-1}$, and thus the particle trajectories emitted under
positive and negative become separated.
This forms a black region in the center of the screen $S_2$ just behind the slit
(Fig.\ref{fig:black}).

Although we succeeded with a model of a field localized near the slit.
We took Gaussian rapidly changing force field of the form
$$
\vect{F}(x,y)=\vect{e}_x~F_0e^{-\sigma x^2}.
$$
It is very stimulating that such setup also produces nontrivial pattern,
since it means that one could start thinking of a physical
nature of this force field.

We have also considered the case of non-point-like Gaussian source distributed
along the vertical line.
This produces smoother picture, which could be obtained by convolving the
pattern on Fig.\ref{fig:histg} with a Gaussian kernel.

\section*{Acknowledgments}

We would like to thank
I would like to thank L.~Accardi, L.~Ballentine, V.~Belavkin, G.~Emch, B.~Dragovich,
R.~Gill, D. Greenberger, K.~Gustafsson, B.~Hiley, G.~t`Hooft, L.V.~Joukovskaya,
A.~Plotnitskii, I.V.~Volovich for stimulating discussions on quantum phenomena.

\appendix

\section*{Numerical Simulation}

Here we provide a {\it Mathematica} program for computation of the
histogram. Please note that although this program is approximately
$10^3$ times slower than a more powerful parallel \textit{C++} program
which we used to actually perform computations it reproduces the same
results (in the simplest case where the computation time allowed us to wait the result)
and could be seen as a detailed description of the
numerical experiment which was performed as a demonstration of the
concept.

\noindent
ComputeTrajectory=Compile[\{\{$\alpha$, \_Real\}\},\\
  With[\{\\
    $d = 5, l = 10, \delta = 0.5, R = 5$,  (* Frame parameters *)\\
    $q = -1$, (* Charge *)\\
    $\tau = 0.025$, (* Discreteness parameter *)\\
    $v = 12$ (* Initial velocity *)\\
  \},\\
  Module[\{\\
  $x = -1.0*d, y = 0.0, Vx = v*Cos[\alpha], Vy = v*Sin[\alpha], Fxv = 0.0, Fyv = 0.0$\},\\
    While[$x \leq l$,\\
$~~~~      Fxv = \mbox{If}[Abs[x]\leqslant\delta, 2 \pi q, 0]$;\\
$~~~~      Fyv = 0$;\\
$~~~~      x += Vx*\tau$;\\
$~~~~      y += Vy*\tau$;\\
$~~~~      Vx += Fxv*\tau$;\\
$~~~~      Vy += Fyv*\tau$;\\
    ];\\
    Module[\{\\
      $xPrev = x - (Vx - Fxv*\tau)\tau$,\\
      $yPrev = y - (Vy - Fyv*\tau)\tau$\},\\
      $(y - yPrev)\frac{\delta - xPrev}{x - xPrev} + yPrev$    ]]]];\\
$ComputeHistogramm[r\_:0.1]:=Module[$\\
$~~\{b,i=1,iPrev,t=\{\},h\},$\\
$~~For[b=tbl[[1]]+r,b<tbl[[-1]],b+=r,$\\
$~~~~      iPrev=i;$\\
$~~~~      While[tbl[[i]]\leq b,i++];$\\
$ ~~~~     t=\{t,h[\{b,i-iPrev\}]\}$\\
$  ~~~~    ];$\\
$   ~~~~ Flatten[t]/.h[a\_]\to a$\\
$  ~~  ];$\\
$Timing[tbl = Table[ComputeTrajectory[\alpha^\circ], {\alpha, -49, 49, 0.01}];]$\\
$hst=ComputeHistogramm[0.025];$\\
$ListPlot[hst,PlotJoined\to True,PlotRange\to All];$\\
$ListContourPlot[Transpose[hint/.{X\_,P\_}\to{P,P}],$\\
$~~~~ContourLines\to False, Contours\to 100,FrameTicks\to False]$

\end{document}